\documentclass[preprint2]{aastex}

\shorttitle{Vega}
\shortauthors{Yoon et al.}

\begin{document}

\newcommand{\degree}{$^{\circ}$}
\newcommand{\ie}{i.\ e.,}
\newcommand{\eg}{e.\ g.,}
\newcommand{\etal}{et al.}
\newcommand{\I}{{\scriptsize I}}
\newcommand{\II}{{\scriptsize II}}
\newcommand{\III}{{\scriptsize III}}
\newcommand{\W}{$ \lambda\, $}
\newcommand{\LL}{$ \lambda\lambda\, $}
\newcommand{\vel}{\,km\,${\rm s^{-1}}\, $}

\title{The effect of rotation on the spectrum of Vega\altaffilmark{8}}

\author{Jinmi Yoon\altaffilmark{1,2}, Deane M. Peterson\altaffilmark{1,3},
Robert J. Zagarello\altaffilmark{4}, J. Thomas
Armstrong\altaffilmark{5,6}, and Thomas Pauls\altaffilmark{5,7}}
\altaffiltext{1}{Department of Physics and Astronomy, Stony Brook University,
    Stony Brook, NY 11794-3800}
\altaffiltext{2}{jyoon@grad.physics.sunysb.edu }
\altaffiltext{3}{dpeterson@astro.sunysb.edu}
\altaffiltext{4}{PCPION, South Setauket, NY 11720-1325; rzagarello@mail.astro.sunysb.edu}
\altaffiltext{5}{Remote Sensing Division, Code 7215, Naval Research Laboratory,
        4555 Overlook Avenue SW, Washington, DC 20375}
\altaffiltext{6}{tom.armstrong@nrl.navy.mil}
\altaffiltext{7}{pauls@nrl.navy.mil}
\altaffiltext{8}{Based 
on spectral data retrieved from the ELODIE archive at Observatoire de
Haute-Provence (OHP)}

\begin{abstract}
The discovery that Vega is a rapidly rotating pole-on star has raised
a number of questions about this fundamental standard, including such
issues as its composition, and in turn its mass and age. We report
here a reanalysis of Vega's composition. A full spectral synthesis
based on the Roche model derived earlier from NPOI interferometry is
used. We find the line shapes in Vega's spectrum to be more complex
than just flat-bottomed, which have been previously reported; profiles
range from slightly self-reversed to simple ``V'' shapes. A high SNR
spectrum, obtained by stacking spectra from the ELODIE archive, shows
excellent agreement with the calculations, provided we add about
10\vel of macroturbulence to the predicted spectra. From the abundance
analysis, we find that Vega shows the peculiar abundance pattern of a
\W Bootis star as previously suggested. We investigate the effects of
rotation on the deduced abundances and show that the dominant
ionization states are only slightly affected compared to analyses
using non-rotating models. We argue that the rapid rotation requires
the star be fully mixed. The composition leads to masses and
particularly ages that are quite different compared to what are
usually assumed.
\end{abstract}

\keywords{line: profiles --- stars: abundances ---
stars: chemically peculiar --- stars: early-type --- 
stars: individual (Vega) --- stars: rotation}

\section{Introduction}
With the announcement of the detection of the interferometric
signature of rapid rotation in Vega
\citep{Peterson2004, Peterson2006b,Aufdenberg2006}, a number of
questions were raised about the fundamental standard. Earlier
suggestions of rapid rotation were based on the high luminosity
\citep{Petrie1964, Gray1988} of the object and the unusual shapes of the weak 
lines in the spectrum \citep{Gulliver1994, Hill2004}. The high
luminosity is immediately explained using Roche models for the figure
of the rotating star, von Zeipel's theorem \citep{Zeipel1924} to
characterize the temperature distribution, and adopting a nearly
pole-on geometry, required by the small line widths \citep{Gray1988}.

But this model, a star rotating near breakup, raises the question of
whether such fundamental issues as Vega's composition, mass, and age,
are accurately known. It has been recognized for some time that Vega
appears metal poor\,\citep{Sadakane1981, Adelman1990}. And although it
has been known since the early 20th century that masses deduced from
luminosity and radius measurements are strongly affected by
composition, recent mass and age determinations have largely assumed
solar composition, the assumption being that sharp lined A stars often
show abundance peculiarities that are assumed due to diffusion and
generally confined to surface layers. The recognition of rotation
velocities approaching breakup renders that assumption unlikely, since
rotation driven circulation is likely to mix the envelope completely
and deeply over times short compared to operable diffusion timescales.

Furthermore, the large surface temperature gradients that would be
associated with high rotation raise a new question: how seriously are
simple, single model atmosphere analyses of the spectrum affected by
the composite nature of the atmosphere? The peculiar line shapes add
to this concern.  A full analysis of the spectrum, or at least
representative spectral features, seems necessary to demonstrate that
we understand the peculiar line profiles and are able to derive
reliable abundances.

This in turn requires high resolution, low noise; spectra comparable to
those used by \citet{Hill2004}. The spectra of Vega available on the
ELODIE archive provide us the necessary resolution and low noise, we
describe those data in \S\,2.  The computation of the synthetic
spectra based on a rotational model are described in \S\,3 and the
deduced abundances and other characteristics we reported in \S\,4,
including the discovery that significant macroturbulence must be
adopted. In \S\,5 we discuss the implications of the abundance profile
and argue that the suggestion that Vega belongs to the \W Bootis class
of objects is probably correct. We note that the effect of rotation on
the line strengths depends strongly on the line considered and propose
a simple resolution to the prediction of large departures from LTE in
the Fe \I\, spectrum that have not been seen in practice. We examine more
closely the issue of rotational mixing and conclude that the abundances
we find here likely represent the material out of which Vega was
formed.  Lastly, we estimate the mass and age of Vega based on this
composition. 

After submitting this manuscript we became aware of a paper 
\citep{Takeda2008} that had been recently accepted for publication 
in this {\it Journal} which undertakes an analysis of their previously
published \citep{Takeda2007} spectra toward understanding Vega's
rotation, much along the lines taken earlier \citep{Gulliver1994,
Hill2004}.  These authors draw a number of conclusions in agreement
with what we find here.  But they also arrive at quite a different
physical model of Vega, concluding in the process that errors were
made in the reductions of the published interferometry.  We will
comment on these results at the appropriate points.

\section{The Observational Data}
The Vega spectra we used are from the ELODIE archive
\citep{Moultaka2004}, which contains high-resolution ($R \sim 42,000$)
echelle spectra from the ELODIE spectrograph obtained at the
Observatoire de Haute-Provence 1.93\,m telescope. The ELODIE data
pipeline automatically extracts the spectra, establishes the
dispersion, and corrects for scattered light. The spectra used here
were obtained between 1996 and 2004.  The wavelength rectified spectra
covering \LL 4000-6800 are provided with 0.05\,\AA \,sampling.
 
\begin{figure*}
\epsscale{1.50}
\plotone{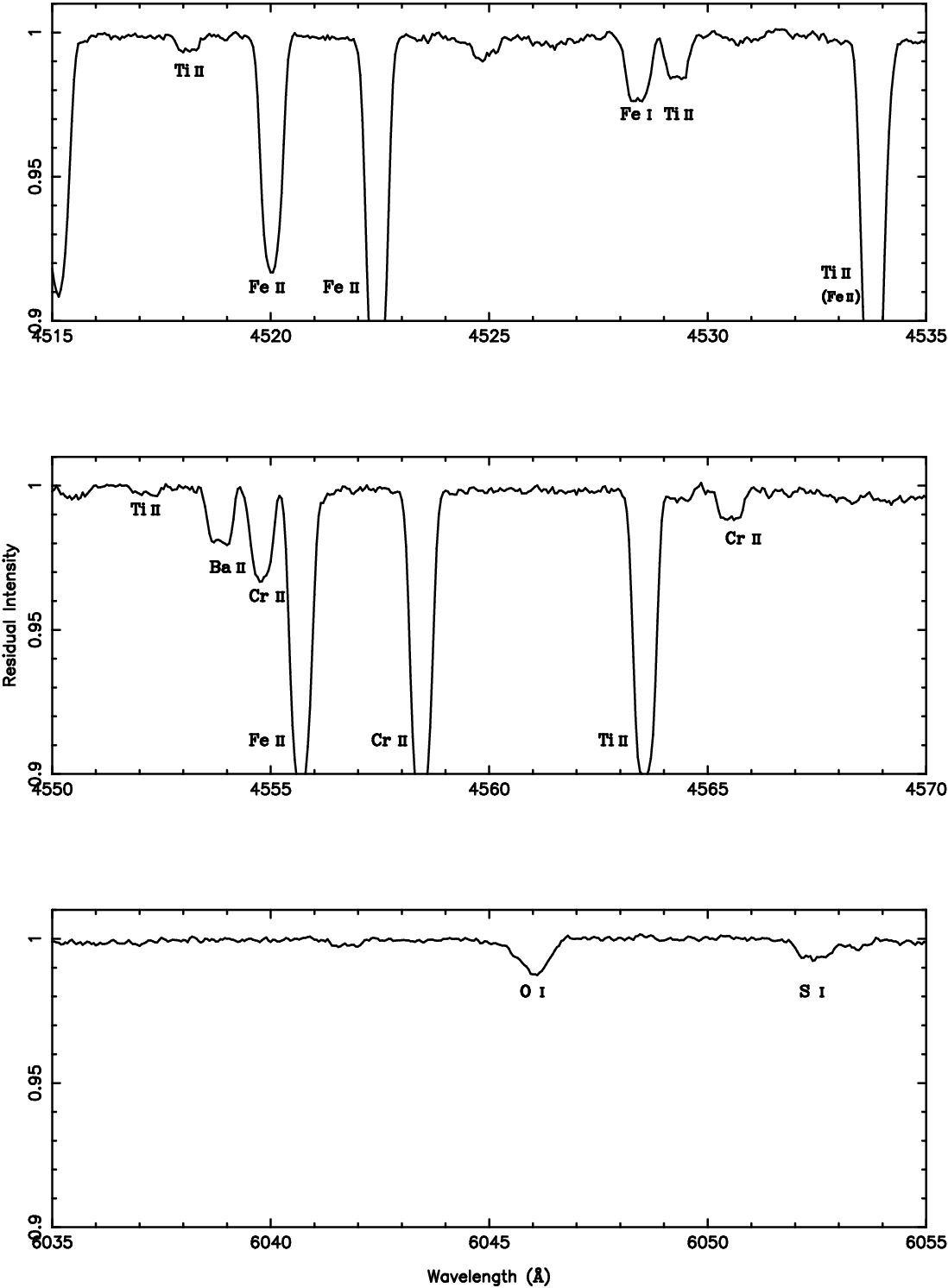}
\caption{Shown are line profiles representative of the range of shapes
encountered for weak lines in the ELODIE spectra of Vega.  The shapes
run from weakly ``self-reversed'' (e.g., Fe~\I\ \W4528 and Ba~\II\ 
\W4554) through flat-bottomed (Cr~\II\ \W4565 and S~\I\ \W6052) to             
``V''-shaped (O~\I\ \W6046).  Where known, blends are indicated in              
parenthesis.  Wavelengths are in the star's rest frame.\label{fig1}}
\end{figure*}

\begin{figure*}
\epsscale{1.50}
\plotone{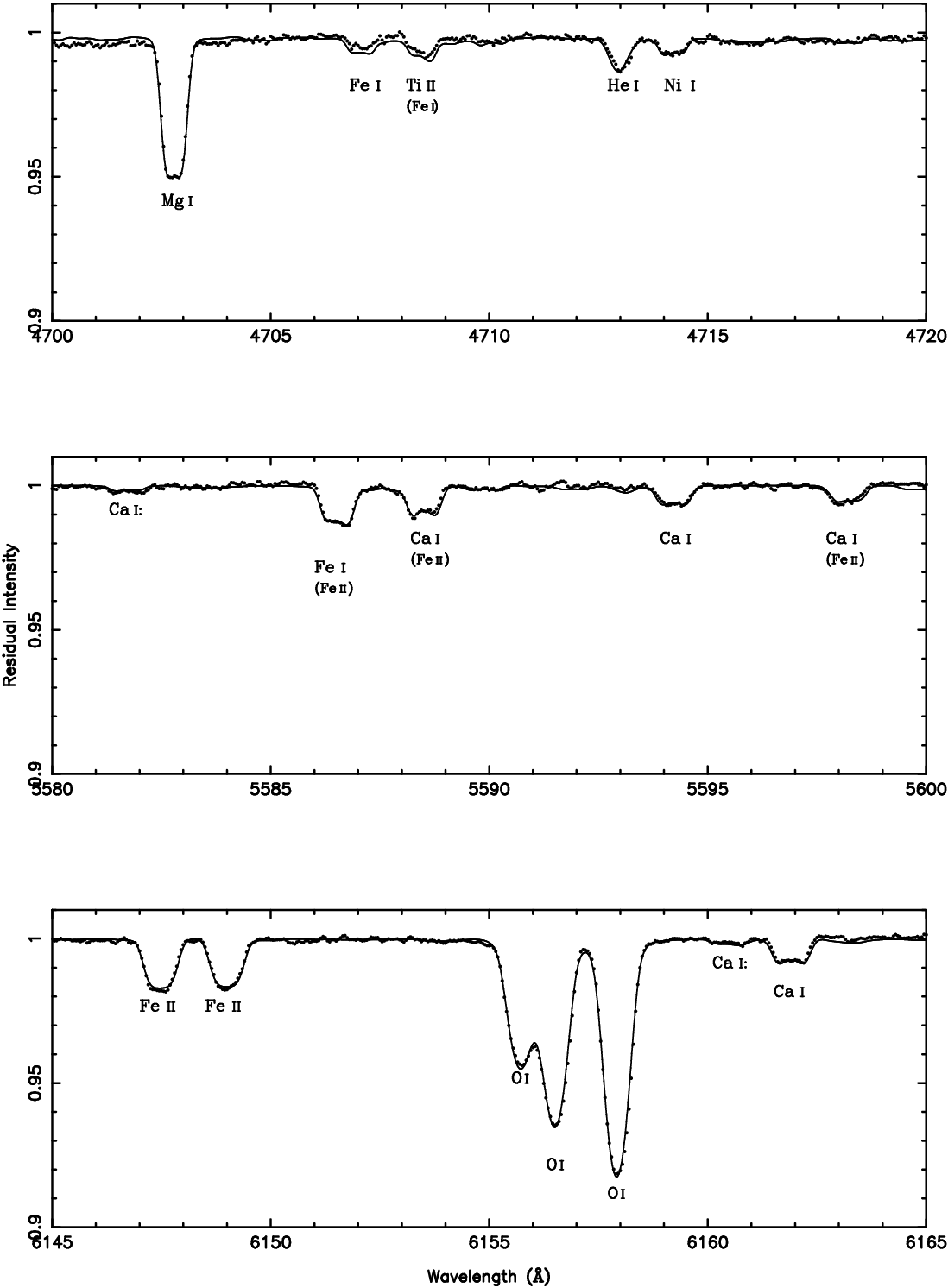}
\caption{ Plotted here are additional segments of spectra (dotted
lines) showing the range of shapes of weak lines, as in
Figure~\ref{fig1}, only now overplotted with the synthetic spectra
(continuous lines). Note particularly He I \W\,4713 which, with an
excitation potential of 21\,eV, is formed in a small region around the
rotational pole and displays the corresponding ``V'' shape.  At the
other extreme Ca~\I\ \W6162 shows the weak double-horned
(``self-reversed'') shape reflecting its very low excitation
potential, 1.9\,eV; it is contributed exclusively by the cooler
equatorial regions.  Other lines showing this behavior are Ti~\II\
\W4708, Fe~\I\ \W5586, and Ca~\I\ \W5588, although all three are
(weakly) blended.  Two iron lines, Fe~\II\ $\lambda\lambda$\,6147 and
6149, at intermediate excitations of 3.9\,eV above the 7.8\,eV
ionization potential of Fe~\I, show the expected flat-bottomed shapes,
although seen against a slight variability in the background
continuum.  The weak Ca~\I\ lines indicated with ``:'' were not
included in the abundance determination. \label{fig2}}
\end{figure*}

Barycenter corrections were required before co-adding the spectra. To
improve the signal-to-noise ratio (SNR) which is 250 for a typical
spectrum, we co-added 49 out of 71 available spectra. In the
process we rejected spectra whose SNR was less than 100 and those
showing noticeable fringing. We also replaced bad pixels whose
residuals in individual spectra were 5 times larger than the typical
noise by interpolating adjacent pixels. The co-added spectra were
converted to a residual intensity scale by normalizing them to the
scale of the synthetic spectra described in \S\,3. The resultant
spectra, segments of which are shown in Figures \ref{fig1} and
\ref{fig2}, were then compared to the synthesized spectra for the
abundance analysis. The SNRs of the co-added spectra were estimated to
range from 750 to 2,200 depending on the spectral regions.

Besides the ELODIE spectra, spectra of comparably high SNR and
resolution of Vega have been obtained at the Dominion Astrophysical
Observatory\,\citep[DAO;][]{Gulliver1994, Hill2004} and the Okayama
Astrophysical Observatory\,\citep[OAO;][]{Takeda2007}. The DAO spectra
(SNR $\sim 3,300$) have not yet been released publicly. However they
are available as part of a graphic
toolkit\footnote{\url{http://www.brandonu.ca/physics/gulliver/ccd\_atlases.html}}
which allows one to examine sections of spectra at high resolution and
identify lines and probable blends. We made extensive use of this tool
during this investigation.\footnote{At the same time R.L. Kurucz
(2007, private communication) provided a high resolution synthesized
spectrum for Vega in the 450--500\,nm region based on a line list and
gf values calibrated to a solar spectrum which proved extremely
useful.}

Also recently published are OAO spectra (SNR from 1,000 to 2,000 on
average) covering \LL 3900-8800. However, these spectra display
emission (\eg \,$\sim$\, \W 4560) and absorption (\eg
\,$\sim$\,\W6060) features and show the head of the Paschen continuum to be
strongly in emission, features not reported elsewhere. So we have
chosen to focus exclusively on the ELODIE data set.

\section{Computations}
We assume Vega can be described by a gravity-darkened Roche spheroid
in solid-body rotation, with a point mass gravitational potential,
showing a temperature distribution varying according to von Zeipel's
theorem \citep{Zeipel1924}, and seen nearly pole-on
\citep[\eg][]{Peterson2006b, Aufdenberg2006}. Because the recent
interferometric measurements taken at the Navy Prototype Optical
Interferometer \citep[NPOI;][]{Armstrong1998} and the Center for High
Angular Resolution Astronomy \citep{tenBrummelaar2005} array yield
closely similar model parameters, we adopt the parameters obtained
from the NPOI data \citep{Peterson2006b} for synthesizing spectra; the
model has a fractional rotation velocity, $\omega =0.926$, a polar
surface gravity of $\log g_{\rm{p}} = 4.074$, a polar effective
temperature of $T_{{\rm p}} = 9988$\,K, an inclination of the
rotational axis to the line of sight $i = 4.54$\,\degree, and a
projected rotational velocity of $v\sin i = 21.7$\vel. For details of
fitting Roche models to the NPOI data see \citet{Peterson2006a}, and
for issues specific to Vega see
\citet{Peterson2006b}, respectively.

To calculate the emergent spectrum we constructed a square $256 \times
256$ grid which contains the apparent disk of the star, calculated the
stellar parameters at the center of each cell that actually fell on
the flattened disk, and computed an emergent flux as a function of
$\lambda$, $\mu$ (cosine of the angle between the local normal and the
line of sight), $T_{{\rm eff}}$, $g_{{\rm eff}}$ (local gravity
reduced by centrifugal force), and projected velocity using the ATLAS9
model atmosphere grid \citep{Castelli2003} and the atomic line data
given in the extensive compilation of \citet{Kurucz1995}. The fluxes
were integrated over the disk to yield the synthetic model
spectrum. In these calculations LTE, hydrostatic equilibrium, and
plane-parallel atmospheres were assumed to represent the star's
surface locally.

A concerns have been raised recently \citep{Aufdenberg2006, Monnier2007}
about the rigorous applicability of the von Zeipel theorem in the parts
of the disk of a rotating star that are rendered cool enough to
generate convection.  We believe the issue is not relevant to Vega.  In
our model the temperature drops to about 7600\,K at the equator, and the
effective gravity in turn decreases to about $\log g \sim 3.5$.  From a
model atmosphere with $T_{\rm eff}=7500$\,K, $\log g=3.5$ we find the
reduced density and in turn increased fraction of hydrogen ionized
compared to the main sequence, substantially decreases the extent of
the convective region and the efficiency of the resulting convection.
Convection carries significant flux only in the range of $1 \leq
\tau_{Rosseland} \leq 30$, well out from the interior where the flux
requirement is established.

\section{Results}
\subsection{Line Shapes}
The abundance analysis was done by adjusting each element abundance
until the model spectra fit the co-added spectra. Since
Vega's lines are sharp and blending is minimal, the process of
adjusting the abundances was straightforward.  Several representative
regions of the co-added ELODIE spectrum are shown in Figures
\ref{fig1} and \ref{fig2}.  Weak lines throughout the spectrum show not
only the flat-bottomed shapes (Cr \II\ \W4565, S \I\ \W6052, and Fe
\II\ \W6147) as noted in recent studies \citep{Gulliver1994,Hill2004} 
but also weakly ``self-reversed'' shapes such as Mg \I\ \W4702 and Ca
\I\ \W6162 and ``V'' shapes such as He\,\I\ \W4713 and O \I\ \W6046.
        
The unusual shapes of the weak lines are strongly correlated with
excitation and ionization potential and can be understood in terms of
how the Boltzmann factors amplify the temperature gradient across the
disk. Since Vega is seen nearly pole on, the center of the apparent
disk is almost exactly at one pole, the hottest point on the star. On
the other hand, the limb is nearly the equator which is not only
2,400\,K cooler than the pole, but the visible gas is actually cooler
still owing to the simple projection effects associated with
limb-darkening. Therefore the bound states responsible for the lines
seen from the light elements such as He \I, O \I, Mg \II, Al \II, and
Si \II\ whose ionization and excitation potentials are quite high are
excited mostly at the axis with zero projected velocity. There is
almost no contribution to the line profiles from the rotationally
shifted equatorial region, resulting in ``V'' shapes. The lower the
excitation potentials the lines have, the more enhanced the
contribution from the more rapidly rotating equatorial regions becomes
and the wider and more square shaped the line profiles get. For the
elements such as Ca \I, Fe \I, and Ba \II\ with the lowest excitation
potentials one sees a mild double-horned shape (``self-reversed'') as
the contribution from the equatorial region completely dominates the
profile. In this sequence the flat-bottomed shape is formed at
intermediate excitation potentials such as those of the lines 
of Cr \II\ and Fe \II. Our synthetic spectra predict well this sequence of
line shapes as shown in Figure \ref{fig2}, where three regions of the
ELODIE spectrum, overplotted with our synthetic spectrum, are
shown. We see that weak Fe \II\ lines tend to have a flat-bottomed
shape while weak Fe \I\ lines show a self-reversed shape. Examination
of data presented by \citet{Gulliver1994} and particularly in
\citet{Hill2004} suggests these shapes are present in their data as
well.

\subsection{Macroturbulence}
\begin{figure*}
\includegraphics[angle=270,scale=.65]{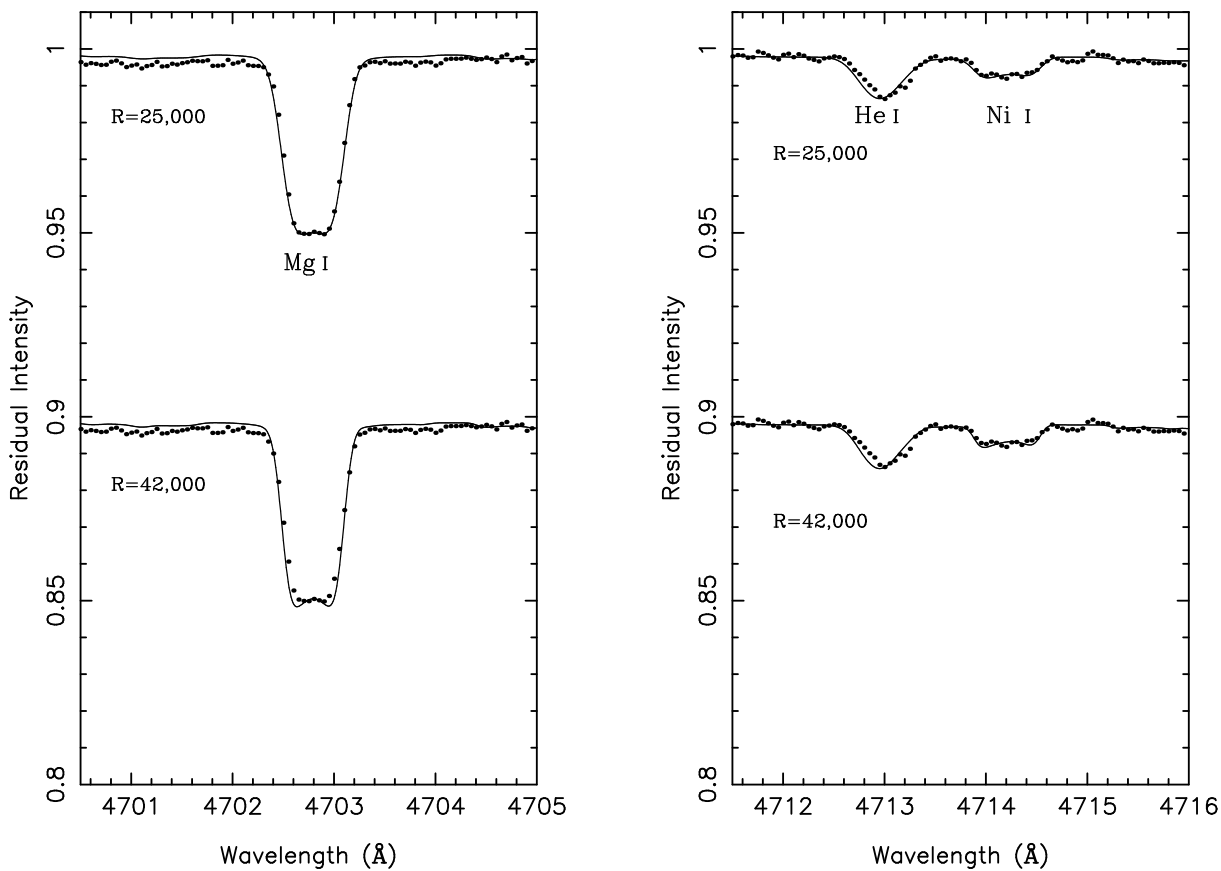}
\caption{These figures plotted as in Figure~\ref{fig2} 
illustrate the need for line broadening in addition to 
rotation and microturbulence (in both panels the lower spectra are
offset by 0.1). Mg~\I\ \W4702 shows the problem most 
clearly although it is also evident in Ni~\I\ \W4714.  The nominal 
ELODIE resolution of 42,000 (assumed to be Gaussian) allows too much 
structure in the steep-sided line profiles.  Reducing the resolution to 
25,000 appears to be required, which we interpret as a contribution of 
about 10 km\,s$^{-1}$ of macroturbulence.  The effect of adding this 
macroturbulence is to improve the fit dramatically in the bottoms of 
the weak, low-excitation lines while causing the line widths to be a 
bit wide.  This suggests the actual projected rotation rate is below 
the adopted 21.7 km\,s$^{-1}$, as was suggested in the initial 
interferometric data reductions reported by \citet{Peterson2006b}. 
Note that high excitation lines like He~\I\ \W4713 are not affected by 
the added macroturbulence. \label{fig3}}
\end{figure*}

In the process of the spectral synthesis, in order to fit the shapes
of the weak line we found we had to reduce the resolution of the
spectra well below the nominal resolution of 42,000 of the ELODIE
spectra, ultimately adopting a resolution of about 25,000 as shown in
Figure \ref{fig3}. We interpret this additional broadening, which was
accomplished by convolving the synthetic spectrum with a Gaussian, as
adding 10\,km\,${\rm s^{-1}}$ of macroturbulence to the nominal ELODIE
resolution (also assumed to be a Gaussian). The effect of this
additional broadening is most noticeable in steep-sided line profiles
(\eg\, Mg \I\ \W4703 and Ni \I\ \W4713), as shown in Figure
\ref{fig3}. As a result the value determined for the macroturbulence
comes from low excitation lines and hence refers more to the
equatorial regions than the polar regions. As might be expected,``V''
shaped lines such as He \I\ \W4713 are insensitive to the
macroturbulence, as also shown in Figure \ref{fig3}. This is a very
interesting result which we discuss at more length below. One caution
is immediately apparent though: line widths might not be reliable
indicators of actual projected velocity, at least for stars seen at
low inclination.

At this point our analysis deviates sharply from the recent
contribution from \citet{Takeda2008}, who seem not to have considered
the possibility of large scale non-thermal line broadening.  That
there could, and even should, be turbulence on large scales in the
atmosphere of Vega seems easy to justify. Even very slow, cm\,s$^{-1}$,
subsurface circulation currents will be magnified by the many order of
magnitude drop in density found in the outer envelope, as required by
the equation of continuity. Add to this a very strong Coriolis force
owing to the rapid rotation and a surface covered with large eddies -
cyclones - is to be expected.  Ignoring this possibility,
\citet{Takeda2008} were forced to adopt a relatively slowly rotating
model, creating a clear conflict with the interferometric measurements
\citep{Peterson2006b,Aufdenberg2006}.

\subsection{Abundance Analysis and Microturbulence}
\begin{figure}
\epsscale{.80}
\plotone{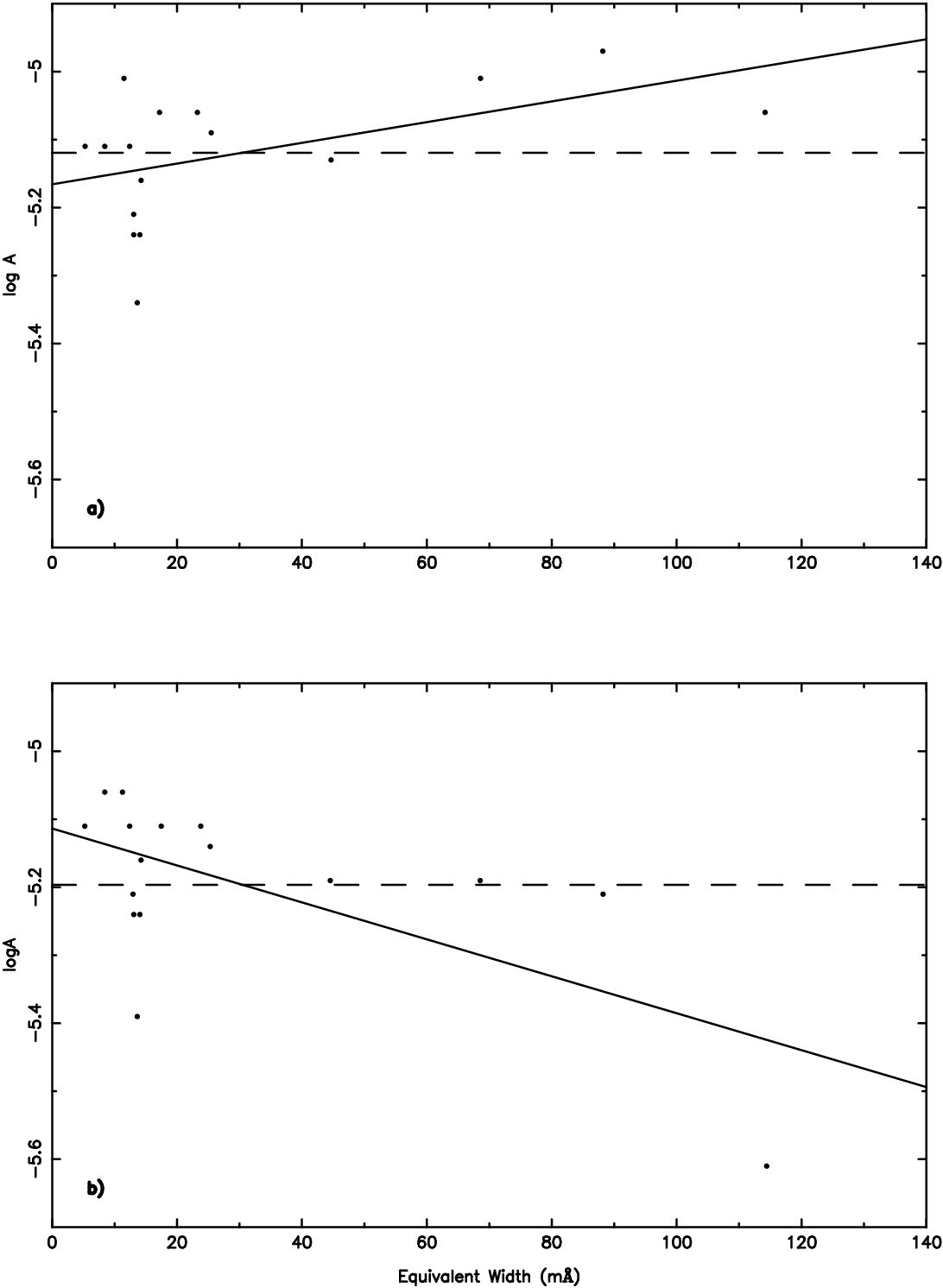}
\caption{ These plots show the derived abundances (data points) versus
equivalent widths of Fe \II\ lines for two different assumed values of
the microturbulence. Panels a) and b) show the derived abundances for
microturbulence values of 2\vel and 4\vel, respectively. The dashed
line shows the unweighted average abundances and the solid line shows
the trend with equivalent width.  We adopt a microturbulence of 2\vel
in our abundance determinations. \label{fig4}}
\end{figure}

As is often the case, we found that it was generally not possible to
find abundances for elements (or even the same ion of an element)
which gave good fits to both strong and weak lines
simultaneously. This is usually taken as a signal that some
microturbulence needs to be introduced. To this end, we determined the
abundances from the Fe \II\ lines for two choices of the
microturbulence, as shown in Figure \ref{fig4}. Here the abundances
are given as the logarithm of the ratio of the number of an element to
that of total elements, $\log\frac{N_{{\rm el}}}{N_{{\rm
tot}}}$. \citet{Castelli1979} and
\citet{Sadakane1981} have previously noted that the influence of
the microturbulence is less important in the visual region for
lines of intermediate strengths about $40 \sim 70 $\,m\AA\,, which
we also found. For Fe \II, which has the widest range of equivalent
widths, we find both the scatter and any trend with equivalent
width are significantly reduced for a microturbulence of about
2\vel which we subsequently adopt.  The O \I\ triplets also support
2\vel (\eg\ Figure \ref{fig2} which shows only the case for
2\vel).

Table \ref{tbl-1} shows the deduced abundances for Vega with
a microturbulence of $\xi_T =$ 2\,km\,s$^{-1}$.  In selecting lines we
eliminated severe blends but included weak blends where we felt
reliable abundances could be obtained. The columns are the laboratory
wavelength, lower excitation potential, equivalent width, $\log gf$,
and the deduced abundance ($\log\frac{N_{{\rm el}}}{N_{{\rm tot}}}$).
Blends we have decided to retain are noted in the last column. The
abundances for elements with only single lines such as Al \II, S \I,
Mn \I, and Ni \I\ must be considered uncertain. Even where there was
no obvious blending, abundances were determined exclusively by
spectral synthesis. Nevertheless, we give equivalent widths for
comparison with recent work; agreement is within 1--2\,m\AA\
typically.  Equivalent widths are missing where lines were not able to
be measured due to ``one-sided'' blends or difficulty in defining the
local continuum level.

Notable in Table \ref{tbl-1} is the discrepancy between Fe \I\ and Fe
\II\ abundances. The abundances of Fe reported by \citet{Adelman1990}
do not show this dramatic lack of balance, and this might be viewed as
supporting the smaller temperature gradient derived by
\citet{Gulliver1994} and \citet{Hill2004}. In contrast with Fe, the
abundances of Mg \I\ and Mg \II\ shown in Table \ref{tbl-1} do not show
similar behavior. We discuss this result further below.

\section{Discussion}
\subsection{How Does Rotation Affect Abundances?}
The main difference between a pole-on rapidly rotating star, as
modeled here, and a classical plane-parallel stellar atmosphere model
is that for the same integrated colors, the rotating model has some
fraction of its surface at both higher and lower local effective
temperatures than the non-rotating model. The expected effect is that
there will be spectral lines in a range of excitation and ionization
energies where the two models give similar results. For Vega this is
the case when the sum of the excitation energy and ionization energy
(for lines of ions such as Fe \II) is about 10\,eV. But for both
higher and lower energy features the expanded range of temperatures
will enhance line strengths, resulting in a decrease in the deduced
abundances from those lines
\citep[described as ``intensification'' by][]{Takeda2008}.  

We see that trend here.  Table \ref{tbl-2} summarizes the results
of two recent analyses of Vega with non-rotating models
\citep{Sadakane1981,Adelman1990} along with the element-by-element
results determined here.  Our results from Mg~\I,
Mg~\II, Al~\II, Si~\II, Ti~\II, Cr~\II, Mn~\I, and Fe~\II\ are about
-0.6 dex below solar, roughly that found by earlier authors.  These
are mostly the dominant ionization stages and thus yield fairly stable
abundances.  Other ions, including C~\I, Ca~\I, Sc~\II, Fe~\I, Ni~\I,
and Ba~\II\ are deficient by about -1.0 dex, or even more, than the
solar abundances (C~\I\ by -0.6), and are depressed by typically
several tenths dex compared to the earlier studies, in this case
because of their relatively small excitation and ionization energies.

On the other side, the He abundance we deduce, $N_{{\rm He}}/N_{{\rm
tot}} =0.072 \pm 0.004$, which is essentially solar (0.078), is
substantially higher than that found by \citet{Adelman1990}, running
counter to expectations.  The result determined here is based on five of
the six lines in the ELODIE spectral window that are expected to be
measurable (\W5875 is heavily involved with atmospheric water vapor
lines), while the helium abundance of \citet{Adelman1990} is based on
\W4471 only. Otherwise, we have no explanation for why we obtain a larger
helium abundance.

In broadest terms, we find that if one can determine abundances from
lines of the dominant ionization stage of an element, the errors
induced by not accounting for rotation are small.  Where lines from
the dominant ionization stage are not accessible (\eg Ba~\III),
one can expect large corrections to be required when standard,
model-atmosphere analyses are applied to objects rotating near
breakup.

One interesting example of the problems that can arise because of
the corrections required between different ionization stages of
the same element, involves the ionization balance between Fe~\I\ and
Fe~\II.  Problems with the Fe~\I/Fe~\II\ ionization balance have
been reported for a wide range of stars
\citep[\eg][]{Gigas1986,Allende1999,Thevenin1999,Johnson2002}. For
Vega, departures from LTE are predicted to produce about 0.3 dex
errors in abundances deduced from Fe~\I\ lines while Fe~\II\ lines are
barely affected \citep{Gigas1986}.  However, these calculations are
difficult owing to the complexity of the atom and the lack of accurate
collision and photoionization cross sections.  For example,
\citet{Pradhan1995} have found that many of the photoionization cross
sections of Fe \I\ are significantly higher than those previously
adopted \citep[\eg][]{Gigas1986} with the possibility that the actual
corrections from departures from LTE are larger still.

The problem with the Fe balance in Vega is confusing since at first
glance, straightforward LTE analyses
\citep{Sadakane1981,Adelman1990} provide apparent agreement between
the abundances deduced from the two ions. This is in contrast to the
sizable departures from LTE required in other similar objects.

However, even though we assume LTE in our analysis here, we also find
a serious iron ionization imbalance amounting to $\sim 0.4$\,dex, but
in the opposite sense of that induced by non-LTE. To understand the
origin of this imbalance we reanalyzed representative lines from the
two iron ionization states and, as a check, from the two magnesium
ions present, using a standard plane-parallel model. We find that
rotation induces an apparent 0.35\,dex error in the Fe~\I/Fe~\II\
ionization balance, while the corresponding effect in the
Mg~\I/Mg~\II\ balance is only about 0.1\,dex.

Thus we reach the amusing conclusion that a simple LTE analysis of
Vega using models which do not account for rotation give a good
ionization balance because of a nearly complete cancellation of the
effects of photoionization-driven departures from LTE in the Fe~\I\
ion, on the one hand, and an enhanced Fe~\I\ line spectrum
contributed by the extensive cool equatorial regions of the model
owing to the favorable viewing geometry, on the other. Note however,
the near balance between these two effects may disappear when one
analyses lines in either the ultraviolet or infrared, owing to the
changing relative contribution of the equatorial regions to the
overall light.

We note that \citet{Takeda2008} have independently commented on the
near cancellation of departures from LTE versus the effects of
rotation in the iron ionization balance.  However, in their
calculation the rotation induced errors are predicted to be about
half those calculated here, owing to the much lower rotation
velocity and the corresponding dramatically reduced temperature
gradient ($\sim900$\,K) in their model.

\subsection{Is Vega a \W Bootis Star?}

Since \citet{Baschek1988} remarked that Vega showed an
abundance pattern similar to the \W Bootis stars, several studies
\citep[\eg][]{Venn1990,Ilijic1998} reported that Vega may be a mild
\W Bootis star.  We confirm that result here.  The abundance
pattern we deduce matches well the main characteristics of the
abundance patterns of \W Bootis stars as summarized, for example by
\citet{Heiter2002}. Elements such as Si, S, Ca, and Sc fall in the
middle of their respective typical ranges while O, Mg, Ti, Cr, Mn, and
Fe are on the high side of normal and Ni and Ba are on the low side.
While most elements fit the \W Bootis abundance pattern well, C and Al
are somewhat out of the reported range. The Al abundance is based on
one line and is not certain, while carbon is off the lower end of the
pattern reported by \citet{Heiter2002}.  However \citet{Paunzen1999},
in an extensive discussion of carbon and oxygen in this group of
objects, find several objects with carbon abundances as low as -0.7 dex
 with respect to the Sun. We conclude that Vega would not be rejected as a \W
Bootis star on the basis of its carbon abundance and the rest of the
abundances determined here are very much in keeping with membership in
this group.

\subsection{Is Vega Well Mixed?}

A presumption, often unstated, about the nature of $\lambda$\ Boo stars
is that the deviations from solar composition are limited to surface
layers \citep[\eg][]{Baschek1992, Holweger1993}, much the same as
has been concluded for the Ap and Am stars which also occupy this part
of the H-R diagram.  But there has always been some concern about that
assumption since unlike the latter groups the $\lambda$\ Boo stars
appear to have a distribution of rotation velocities similar to normal
stars \citep[\eg][]{Holweger1993}.

We argue here that since Vega is rotating at a significant fraction of
breakup and yet displays fairly typical \W Boo characteristics, it is
unlikely that these composition anomalies are limited to the surface;
more likely, Vega is well mixed. The literature on rotationally induced
mixing has generally focused on the surface layers and the question of
whether the Ap and Am phenomena could be understood as due to diffusive
separation \citep[\eg][]{Charbonneau1993} and not on how fast an 
inhomogeneity introduced on the surface would be mixed throughout the 
envelope.  

However, recent efforts to include the effects of rotation in
evolutionary calculations of massive stars \citep{Meynet1997} have led
to an examination of how inhomogeneities will be redistributed through
a star \citep{Talon1997,Ekstrom2008}, suggesting that extensive mixing
is to be expected. In fact, at the highest velocities in models down to
3 $M_{\odot}$, the lowest mass examined, the mixing is predicted to be
so deep there is the possibility that some of the nuclear products from
the CNO burning region might be mixed to the surface.

This is an interesting possibility, given the low carbon abundance we
have found.  From this point of view, missing is an estimate of the
nitrogen abundance, the lines of which are out of the ELODIE spectral
range.  However measurements of nitrogen line equivalent widths have
been reported elsewhere.  To fill in the abundance of this important
nuclide taking full account of the effects of rotation, we have
calculated the abundances for the nitrogen equivalent widths reported
in \citet{Venn1990} for \W7442.28 ($\log {N/N_{{\rm tot}}}=-4.05$,
$[N/N_{{\rm tot}}]= +0.07$) and \W7468.29 ($\log N/N_{{\rm tot}}=-4.02$, 
$[N/N_{{\rm tot}}]=+0.1$) (\W7423 appears to be blended and we exclude
it here), finding values quite close to those deduced by
\citet{Venn1990} at about 0.085 dex above solar.  This is an
intriguing result.  Although it is difficult to know what ``normal''
is in this star, normalizing to oxygen gives [N/O] $\sim$ +0.2 and
[C/O] $\sim$ -0.5, which may very well indicate that some CN cycle
processed material has been mixed into the envelope of the star.  In
this regard we note that Vega represents a rather unique object; a few
other \W\ Boo objects have projected velocities in the vicinity of
200\,km\,s$^{-1}$, but Vega is the one object known to rotate as fast
as 275\,km\,s$^{-1}$, less than 10\% from breakup in terms of angular
velocity.  However, without a better understanding of the composition
of the material Vega started with, or other supporting information, we
must leave this as just an intriguing possibility.

In summary, we believe a fairly strong case can be made for the outer
layers of Vega being well mixed, possibly even down to the edge of its
nuclear burning core.  If this is so then we are looking at about 2
$M_{\odot}$ of material of highly unusual composition in an object that
is much too young to display such extreme depletion in heavy elements.
In this case the various mechanisms put forth to explain the $\lambda$\
Boo phenomena that rely on its being limited to the superficial layers
\citep[\eg][]{Kamp2002} seem excluded for Vega.  Some form of dust -
gas separation, such as suggested by \citet{Venn1990} or
\citet{Holweger1992}, may be involved but if so the mechanism likely
must work at the time of Vega's formation since so much mass is
involved.  

\subsection{Determination of the Age and Mass of Vega}
We estimate the mass and age of Vega by locating its measured
luminosity and polar radius in an appropriate evolution grid, 
as described in \citet{Peterson2006b}. The interior models we adopt
are from the BASTI database\footnote{www.te.astro.it/BASTI/index.php} 
\citep[][and references therein]{Pietrinferni2006} which include 
evolutionary calculations using scaled solar and alpha-enhanced
compositions for stellar masses up to 2.4\,$M_\odot$. The composition
found here is not a perfect fit to either of the those mixtures, but
the large enhancement of oxygen is about the same compared to the
heavy metals as the alpha-enhanced mixture adopted there. Missing are
the other alpha-rich elements at the enhanced levels, but given the
dominance of oxygen even among these nuclei, that grid should give
results more than adequate. To quantify how much the mismatch in the
details of the distribution of abundances might affect the estimate we
also calculate the mass and age using the scaled solar grid. In both
cases, the heavy element fraction used is ${\rm
Z}=0.0093^{+0.0006}_{-0.0005}$ as calculated from Table
\ref{tbl-2} and assuming $[N_{{\rm el}}/N_{{\rm tot}}] = -0.7$ 
for abundances not obtained here.

For the alpha-enhanced composition we obtain $2.09\pm 0.03 M_{\odot}$
and $536\pm29$\,Myr for the mass and age. The simple scaled
solar abundances in turn yield $2.14 M_\odot$ and 541\,Myr and
since the alpha-enhanced models are a much closer match to Vega's
composition it is clear the errors introduced by the slight
mismatch are small compared to the other uncertainties.

As described in \citet{Peterson2006b}, whether Vega is solar
composition throughout or the derived abundances represent the actual
overall composition, results in quite different estimates for the
star's mass and age.  Most previous authors have assumed an underlying
solar composition yielding estimates of $2.3 M_{\odot}$ for the mass
and an age in the neighborhood of 360\,Myr.  Since there is a distinct
possibility that the composition we have derived applies to the star as
a whole, Vega's estimated mass may be reduced and its implied age
increased substantially.  One immediate consequence of this is a
growing clash with the properties of the so-called ``Castor moving
group'' \citep{BarradoyNavascues1998}, which includes Castor ($\alpha$
Gem), Fomalhaut ($\alpha$ PsA) and Alderamin ($\alpha$ Cep), in
addition to Vega, and whose members are estimated to have an age of
$200\pm100$\,Myr.  Even with an assumed solar composition Vega's age
was not a comfortable fit for inclusion in this group.  The increased
age we propose would make it an unlikely member.

\acknowledgments
This research was supported in part by a grant from the Naval Research
Laboratory to D. M. P. and in part by NSF grant 06-07612 to Dr. Michal
Simon. We also thank Dr. Robert L. Kurucz for extensive discussions.

\clearpage
\begin{deluxetable}{lrrrrl}
\tabletypesize{\scriptsize}
\tablecaption{The Abundance Analysis of Vega\label{tbl-1}}
\tablewidth{0pt}

\tablehead{
\colhead{$\lambda$} & \colhead{EP} &
\colhead{$w_{\lambda}$} &
\colhead{$\log gf$} &\colhead{log $\frac{N_{{\rm el}}}{N_{{\rm
tot}}}$\tablenotemark{a}} & \colhead{Blends} \\
\colhead{(\AA)} & \colhead{(${\rm cm^{-1}}$)}&
\colhead{( m\AA)} &
\colhead{} &\colhead{} & \colhead{}
}
\startdata

\multicolumn{6}{l}{He {\tiny I} ($\frac{N_{{\rm He}}}{N_{{\rm tot}}}$ =0.072 $\pm$ 0.004 )}\\
4471.498 &  169087.008 & \nodata&    0.052 &  0.070 &\\  
4713.139 &  169086.864 &  5 &   -1.233 &  0.078 &Fe {\tiny II} \W4713.193 \\     
4921.931 &  171135.000 &  8 &   -0.435 &  0.060 & \\
5015.678 &  166277.546 &\nodata &   -0.820 &  0.078 &  Fe {\tiny II} \W5015.755\\
6678.154 &  171135.000 &  5 &    0.329 &  0.070 & Fe {\tiny II} \W6677.306 \\
\\
\multicolumn{6}{l}{C {\tiny I} ($\log\frac{N_{{\rm C}}}{N_{{\rm tot}}}$ = -4.14 $\pm$0.04, [$N_{\rm{C}}$/$N_{\rm{tot}}$]\tablenotemark{b} = -0.62)}\\
4770.021 &  60352.639 &  7  &   -2.052 & -4.16& \\ 
4771.730 &  60393.148 & 25  &   -1.488 & -4.16&\\
4775.889 &  60393.148 &  7  &   -2.013 & -4.16&\\
4932.050 &  61981.818 & 16  &   -1.574 & -4.06&\\
\\
\multicolumn{6}{l}{O {\tiny I} ($\log\frac{N_{{\rm O}}}{N_{{\rm tot}}}$ = -3.32 $\pm$ 0.04, [$N_{\rm{O}}/N_{\rm{tot}}$] = -0.11)}\\

5329.099 &   86625.757 & 34\tablenotemark{c}  &   -1.730 & -3.31&\\
5329.690 &   86627.778 & \nodata&   -1.410 & -3.31& \\
5330.741 &   86631.454 & 24 &   -1.120 & -3.31&  \\
6046.438 &   88631.146 & 10 &   -1.675 & -3.26& \\
6155.971 &   86625.757 & 77\tablenotemark{d}  &   -1.051 & -3.36& \\  
6156.778 &   86627.778 & \nodata &   -0.731 & -3.36&  \\
6158.187 &   86631.454 & 59  &   -0.441 & -3.36& \\
\\
\multicolumn{6}{l}{Mg {\tiny I} ($\log\frac{N_{{\rm Mg}}}{N_{{\rm tot}}}$ = -5.12 $\pm$ 0.05, [$N_{\rm{Mg}}/N_{\rm{tot}}$] = -0.66)}\\

4702.991 &   35051.264 & 29  &   -0.666 & -5.06& \\
5167.321 &   21850.405 & 81  &   -1.030 &-5.06 & Fe {\tiny I} \W5167.488\\
5172.684 &   21870.464 &102  &   -0.402 & -5.16 &\\
5183.604 &   21911.178 &119  &   -0.180 & -5.16&\\
5528.405 &   35051.264 & 28  &   -0.620 & -5.16 &\\          
\\
\multicolumn{6}{l}{Mg {\tiny II} ($\log\frac{N_{{\rm Mg}}}{N_{{\rm tot}}}$ = -5.06 $\pm$ 0.04, [$N_{{\rm Mg}}/N_{\rm{tot}}$] = -0.6)}\\

4427.994 &   80619.500 & \nodata &   -1.210 & -5.06 &  \\
4433.988 &   80650.020 &\nodata  &   -0.910 & -5.11 &Fe {\tiny I} \W4433.782\\ 
4481.126 &   71490.190 &\nodata  &    0.740 & -5.01 &\\       
\\
\multicolumn{6}{l}{Al {\tiny II} ($\log\frac{N_{{\rm Al}}}{N_{{\rm tot}}}$ = -6.22, [$N_{{\rm Al}}/N_{\rm{tot}}$] =  -0.65)} \\

4663.046 &   85481.350 &\nodata  &   -0.284 & -6.22 &\\  
\\
\multicolumn{6}{l}{Si {\tiny II} ($\log\frac{N_{{\rm Si}}}{N_{{\rm tot}}}$ = -5.15 $\pm$ 0.05, [$N_{{\rm Si}}/N_{\rm{tot}}$] = -0.66)}\\

4128.054 &   79338.500 & 32 &    0.316 & -5.19 & Mn {\tiny II} \W4128.129\\
4130.872 &   79355.020 & 54 &   -0.824 & -5.19 & \\ 
5055.984 &   81251.320 & 60 &    0.593 & -5.19 & \\
6347.109 &   65500.470 &118 &    0.297 &-5.09 & Mg {\tiny II} \W6346.742 \\
         &             &    &          &      & Mg {\tiny II} \W6346.964\\
6371.371 &   65500.470 & 82 &   -0.003 & -5.09 &\\   
\\
\multicolumn{6}{l}{S {\tiny I} ($\log\frac{N_{{\rm S}}}{N_{{\rm
tot}}}$ = -5.01, [$N_{{\rm S}}/N_{\rm{tot}}$] = -0.3)}\\

6052.674 &   63475.051 &  7 &   -0.740 & -5.01 &\\
\\
\multicolumn{6}{l}{Ca {\tiny I} ($\log\frac{N_{{\rm Ca}}}{N_{{\rm tot}}}$ = -6.72 $\pm$ 0.12, [$N_{{\rm Ca}}/N_{\rm{tot}}$] = -1.04)}\\

4226.728 &       0.000 &\nodata  &    0.243 & -6.73 &\\   
4434.957 &   15210.063 &\nodata  &   -0.029 & -6.73&\\
4585.865 &   20371.000 &  1  &   -0.386 & -6.68 &\\ 
5588.749 &   20371.000 &  1  &    0.210 & -6.63 &\\   
5594.462 &   20349.260 &  5  &   -0.050 & -6.63 &\\  
5598.480 &   20335.360 &  4  &   -0.220 & -6.63 &Fe {\tiny I} \W5598.287\\
6162.173 &   15315.943 &  9  &    0.100 & -6.98 &\\   
\\
\multicolumn{6}{l}{Sc {\tiny II}  ($\log\frac{N_{{\rm Sc}}}{N_{{\rm tot}}}$ = -9.97 $\pm$ 0.05, [$N_{{\rm Sc}}/N_{\rm{tot}}$] = -1.1)}\\

4246.822 &    2540.950 &  5  &    0.320 &-10.02 &\\
5526.79  &   14261.320 &  9  &    0.130 & -9.92 &\\  
\\

\multicolumn{6}{l}{Ti {\tiny II} ($\log\frac{N_{{\rm Ti}}}{N_{{\rm tot}}}$ = -7.65 $\pm$ 0.09, [$N_{{\rm Ti}}/N_{\rm{tot}}$] = -0.63)}\\
4468.507 &    9118.260 & 70  &   -0.600 & -7.82&\\
4529.474 &   12676.970 &  9  &   -1.830 & -7.69&\\ 
4563.761 &    9850.900 & 57  &   -1.010 & -7.51& \\
4589.958 &    9975.920 & 16  &   -1.790 & -7.61&Cr {\tiny II} \W4589.901\\
4708.665 &    9975.920 &  3  &   -2.410 & -7.69 & \\     
4779.985 &   16515.860 & 12  &   -1.420 & -7.59 &\\  
4805.085 &   16625.110 & 21  &   -1.100 & -7.59 &\\        
5336.771 &   12758.110 & 12  &   -1.700 & -7.72 &\\  
\\

\multicolumn{6}{l}{Cr {\tiny II} ($\log\frac{N_{{\rm Cr}}}{N_{{\rm tot}}}$ = -6.91 $\pm$ 0.1, [$N_{{\rm Cr}}/N_{\rm{tot}}$] = - 0.54)}\\
4252.632 &   31117.390 &  6  &   -2.018 & -6.97&\\
4261.847 &   25033.700 & 18  &   -3.004 & -6.92 &Cr {\tiny II} \W4261.913\\
4554.988 &   32836.680 & 20  &   -1.430 & -6.87 &  \\
4558.650 &   32854.310 & 61  &   -0.660 & -6.87 &  \\ 
4565.740 &   32603.400 &  7  &   -1.910 & -7.07 &  \\ 
4588.199 &   32836.680 & 48  &   -0.830 & -6.87 &  \\
4592.049 &   32854.950 & 18  &   -1.420 & -6.87 &  \\
4616.629 &   32844.760 & 16  &   -1.530 & -6.87 &\\  
4618.803 &   32854.950 & 36  &   -1.070 & -6.87 &\\  
4634.070 &   32844.760 & 29  &   -1.220 & -6.82 &\\   
4812.337 &   31168.580 &  6  &   -1.930 & -7.07 &\\
4824.127 &   31219.350 & 39  &   -1.220 & -6.72&\\
5334.869 &   32844.760 & 10  &   -1.562 & -7.07 &\\  
\\    	 

\multicolumn{6}{l}{Mn {\tiny I} ($\log\frac{N_{{\rm Mn}}}{N_{{\rm tot}}}$ = -7.45, [$N_{{\rm Mn}}/N_{\rm{tot}}$] = -0.8)}\\
4783.405 &    18531.663 &  2 &  0.042 & -7.45&\\   
\\

\multicolumn{6}{l}{Fe {\tiny I} ($\log\frac{N_{{\rm Fe}}}{N_{{\rm tot}}}$ = - 5.51 $\pm$ 0.1, [$N_{{\rm Fe}}/N_{\rm{tot}}$] =-0.97)}\\
4132.058 &   12968.553 & 29 &   -0.650 & -5.54 &Fe {\tiny I} \W4131.935\\
         &             &    &          &       &Fe {\tiny I} \W4131.971\\
4134.677 &   22838.321 &  8 &   -0.490 & -5.54 &Fe {\tiny I} \W4134.42	\\
4136.998 &   27543.001 &  4 &   -0.540 & -5.54 &\\	 
4250.119 &   19912.494 & 18 &   -0.405 & -5.49 &Fe {\tiny II} \W4250.437\\
4250.787 &   12560.933 & 27 &   -0.710 & -5.49 &Fe {\tiny II} \W4250.437\\
4260.474 &   19350.890 & 36 &   -0.020 & -5.39 &   \\
4466.551 &   22838.321 &  9 &   -0.590 & -5.36 &   \\
4476.019 &   22946.814 &  8 &   -0.570 & -5.79 &Fe {\tiny I} \W4476.076\\
4528.614 &   17550.180 & 15 &   -1.072 & -5.51 &   \\
4918.994 &   23110.937 & 16 &   -0.640 & -5.51 &   Fe {\tiny I} \W4918.954 \\
4920.502 &   22845.867 & 27 &   -3.955 & -5.51 &   Cr {\tiny II} \W4920.23 \\
5324.179 &   25899.987 & 12 &   -0.240 & -5.49 &   \\
5586.756 &   27166.818 & 11 &   -0.210 & -5.59 &  Fe {\tiny II} \W5587.114\\
5615.644 &   26874.548 & 15 &   -0.140 & -5.44 &  \\
\\

\multicolumn{6}{l}{Fe {\tiny II} ($\log \frac{N_{{\rm Fe}}}{N_{{\rm tot}}}$ = -5.12 $\pm$ 0.09, [$N_{{\rm Fe}}/N_{\rm{tot}}$] = -0.58)}\\
4258.154 &   21812.055 & 14 &  -0.467 & -5.34 & Fe {\tiny II} \W4258.34\\
4520.224 &   22637.205 & 45 &  -2.990 & -5.13 &\\			    
4522.634 &   22939.358 & 69 &  -2.700 & -5.01 &\\			    
4576.340 &   22939.358 & 23 &  -3.390 & -5.06 &  \\			    
4582.835 &   22939.358 & 17 &  -3.570 & -5.06 &  \\			    
4583.837 &   22637.205 & 88 &  -2.490 & -4.97 &  Fe {\tiny II} \W4583.999\\
4596.015 &   50212.826 & \nodata&  -2.057 & -5.21 &  Fe {\tiny II} \W4595.682\\
4620.521 &   22810.357 & 14 &  -3.650 & -5.16 &  \\			    
4635.316 &   48039.090 & 13 &  -1.650 & -5.21 &  \\			    
4656.981 &   23317.633 & 12 &  -3.950 & -5.11 & Ti {\tiny II} \W4657.206  \\
4663.708 &   23317.633 &  5 &  -4.145 & -5.11 & \\   		    
4666.758 &   22810.357 & 11 &  -3.700 & -5.01 & \\ 			    
4670.182 &   20830.582 &  8 &  -4.350 & -5.11 &  Sc {\tiny II} \W4670.407\\
4923.927 &   23317.633 &114 &  -1.820 & -5.06 & \\  			    
5534.847 &   26170.181 & 25 &  -2.930 & -5.09 &  \\  		    
6147.741 &   31364.440 & 14 &  -2.721 & -5.24 &    \\		    
6149.258 &   31368.450 & 13 &  -2.724 & -5.24 & \\			    
\\
\multicolumn{6}{l}{Ni {\tiny I} ($\log\frac{N_{{\rm Ni}}}{N_{{\rm tot}}}$ =  -6.79, [$N_{{\rm Ni}}/N_{\rm{tot}}$] = -1.0)}\\
4714.417 &    27260.894 &  3  &    0.160 & -6.79 &\\   
\\

\multicolumn{6}{l}{Ba {\tiny II} ($\log \frac{N_{{\rm Ba}}}{N_{{\rm tot}}}$ =  -11.21, [$N_{{\rm Ba}}/N_{\rm{tot}}$] = -1.3)}\\
4554.029 &       0.000 & 13 &   0.430 &-11.21 & \\
4934.076 &       0.000 &  7 &  -0.150 &-11.21 & Fe {\tiny I} \W4934.005\\
\\

\enddata
\tablenotetext{a}{For helium abundance, $\frac{N_{{\rm He}}}{N_{{\rm tot}}}$ }
\tablenotetext{b}{[$N_{{\rm el}}/N_{\rm{tot}}$] = $\log\frac{N_{{\rm el}}}{N_{{\rm
tot}}} - \log{(\frac{N_{{\rm el}}}{N_{{\rm tot}}})}_{\odot}$ }
\tablenotetext{c}{The equivalent width is for the blend with O {\tiny I} \W 5629.690.}
\tablenotetext{d}{The equivalent width is for the blend with O {\tiny I} \W 6156.778. }

\end{deluxetable}
\clearpage

\begin{table}
\begin{center}
\caption{Comparisons with the previous abundance studies\label{tbl-2}}
\begin{tabular}{lcccccc}
\tableline\tableline
Atomic && \multicolumn{2}{c}{log M/H } & &$\log\frac{N_{{\rm el }}}{N_{{\rm
tot}}}\tablenotemark{c}$ & [$N_{{\rm el}}/N_{\rm{tot}}$]\tablenotemark{d}\\
\cline{3-4}\cline{6-7}
Species  & &SN\tablenotemark{a} &AG\tablenotemark{b} & & \multicolumn{2}{c}{This work} \\
\tableline
He \I  &&\nodata&-1.52  && -1.14 &-0.04   \\
C \I   &&\nodata& -3.81 && -4.14 & -0.62 \\
N \I\,\tablenotemark{e}&&\nodata&\nodata &&-4.53 & +0.09\\ 
O \I   &&\nodata&\nodata&& -3.32 & -0.11 \\
Mg \I  &&-4.61  & -5.07 &&-5.12  & -0.66 \\
Mg \II &&-4.96  &-5.11  &&-5.06  & -0.60 \\
Al \II &&\nodata&-6.33  && -6.22 & -0.65 \\
Si \II &&\nodata&\nodata&& -5.15 & -0.66 \\
S \I   &&\nodata&\nodata&&-5.01  & -0.30 \\
Ca \I  &&-6.11  &-6.21  &&-6.72  & -1.04 \\
Sc \II &&-9.42  &-9.62  && -9.97 & -1.10 \\
Ti \II &&-7.31  &-7.47  && -7.65 & -0.63  \\
Cr \II &&-6.90  &-6.76  && -6.91 &-0.54  \\
Mn \I  &&-6.87  &-7.16  &&-7.45  & -0.80 \\
Fe \I  &&-5.09  &-5.05  && -5.51 & -0.97 \\
Fe \II &&-5.09  &-5.12  && -5.12 & -0.61 \\
Ni \I  &&-5.94  &-6.38  && -6.79 & -1.00  \\
Ba \II &&-10.25 &-10.58 &&-11.21 & -1.30 \\
\tableline
\end{tabular}
\tablecomments{The definition of the abundances we use differs from 
that adopted by SN and AG. For the helium abundance found here, the SN
 and AG abundances will be systematically larger than ours by 0.03 dex.}
\tablenotetext{a}{\citet{Sadakane1981}}
\tablenotetext{b}{\citet{Adelman1990}}
\tablenotetext{c}{Abundances from Table \ref{tbl-1}}
\tablenotetext{d}{Solar abundances have been taken from \citet{Grevesse1998}}
\tablenotetext{e}{Abundance based on \citet{Venn1990} equivalent widths}
\end{center}
\end{table}

\end{document}